\documentclass[prl,reprint]{revtex4-1}
\usepackage{graphicx}
\usepackage{amsmath,amssymb}
\usepackage{upgreek}
\usepackage{enumitem}
\usepackage{url}
\usepackage{bm}
\usepackage{color}

\newcommand{\kB}{k_{\mathrm{B}}}
\newcommand{\kT}{\kB T}

\newcommand{\um}{\upmu\mathrm{m}}

\newcommand{\phorco}{\Lambda_{\mathrm{PH}}}

\newcommand{\dpcyto}{D_{\mathrm{P}}^{(c)}}

\newcommand{\cATP}{c_{\mathrm{ATP}}}
\newcommand{\DATP}{D_{\mathrm{ATP}}}

\newcommand{\vadvect}{v_{\mathrm{adv}}}

\newcommand{\gradlength}{l_{\mathrm{G}}}

\newcommand{\latin}[1]{{\itshape #1}}

\newcommand{\ie}{\latin{i.\,e.}}
\newcommand{\etal}{\latin{et al.}}

\newcommand{\Eqref}[1]{Eq.~\eqref{#1}}

\newcommand{\Figref}[1]{Fig.~\ref{#1}}

\begin{document}

\title{Diffusiophoresis in Cells: a General Non-Equilibrium, Non-Motor
Mechanism for the Metabolism-Dependent Transport of Particles in Cells}

\author{Richard P. Sear}

\affiliation{Department of Physics, University of Surrey, Guildford,
  GU2 7XH, UK}

\email{r.sear@surrey.ac.uk}
 \homepage{https://richardsear.me/}

\begin{abstract}
The more we learn about the cytoplasm
of cells, the more we realise that
the cytoplasm is not uniform but instead
is highly inhomogeneous.
In any inhomogeneous solution, there are concentration gradients, and particles move either up or down these gradients due to a mechanism called diffusiophoresis. I estimate that inside metabolically active cells, the dynamics  of particles can be strongly accelerated by diffusiophoresis, provided that they are at least tens of nanometres across. The dynamics of smaller objects, such as single proteins are largely unaffected.
\end{abstract}

\maketitle

\begin{figure}[b!]
 \includegraphics[width=8.7cm]{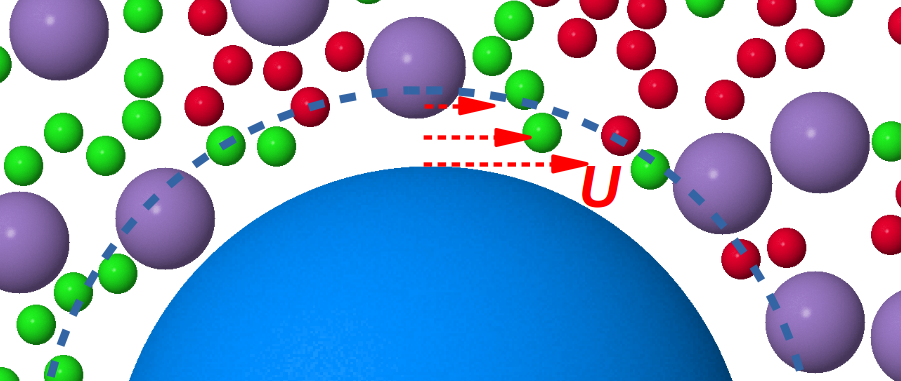}
\caption{Schematic of a particle (blue),
immersed in a cytoplasm with gradients in the concentrations of two metabolites (red and green). Proteins are magenta. Small ions are not shown. The dotted line indicates the approximate extent of the particle/cytoplasm interface, where slip occurs, creating a gradient in velocity, and hence the diffusiophoretic slip velocity ${\bf U}$. 
\label{snapshot}}
\end{figure}

The cytoplasm of cells is far from thermodynamic
equilibrium, and far from uniform \cite{luby99,agutter00,shin17cliff,woodruff17}.
Here, I consider the effect of concentration
gradients on the motion of large particles in the cytoplasm.
Large means tens of nanometres and above, so an example would be a large protein assembly. In the cytoplasm,
particles and molecules are not diffusing alone in a dilute solution, but are moving in a concentrated, active and non-uniform
mixture of proteins, nucleic acids,
metabolites such as ATP, small ions such as potassium, etc. A schematic
of a particle in the cytoplasm, is
shown in \Figref{snapshot}.

It is well known in the fields of colloids
\cite{And86,And89,ruckenstein81,Bra11,sear17,PAN+15,SUS+16,marbach17,yoshida17,bocquet10,FMH+14,VGG+16,shin18,prieve18} and of liquid
mixtures \cite{PAN+15,McAfee201346}, that
particles of one species will move
in response to a gradient in the
concentration of another species. In colloids this is called diffusiophoresis. Diffusiophoresis is typically defined \cite{And86,And89,ruckenstein81,Bra11,PAN+15} as the motion of a larger particle immersed in concentration gradients of smaller molecules, when both are in a solvent (such as water).
Although often difficult to measure, there are clearly
gradients inside metabolically active cells.
So there must be diffusiophoresis
occurring in cells, the question is:
Does diffusiophoresis make a significant contribution
to the transport of some species?
Here, I determine that the answer to this question is probably yes for particles at least tens of nanometres or more across, but no for individual protein molecules. 

I start with the standard Brownian-dynamics approximation for the position of a particle, ${\bf r}(t)$ \cite{allen17}.
With this approximation, we can write the change in position over the time interval $t$ to $t+\delta t$, as \cite{allen17},
\begin{eqnarray}
{\bf r}(t+\delta t) &=& {\bf r}(t)
+\left(2\dpcyto\delta t\right)^{1/2}{\bf \rho}+
\nonumber \\
&&\left[{\bf \vadvect}+{\bf U}\right]\delta t 
+\frac{\dpcyto}{kT}{\bf f}\delta t  ~~~~~
\label{bd}
\end{eqnarray}
This equation includes four possible transport mechanisms for the particle.
The second term on the right-hand side is the conventional thermal
diffusion term. There $\dpcyto$ is the diffusion constant
for thermal diffusion in the cytoplasm, and
${\bf \rho}$ is a vector of random numbers
drawn from a Gaussian distribution of mean zero, and standard deviation one.
The physics of this term is that
the particle is constantly being
bombarded by the surrounding
molecules, due to their thermal
energy. This tends to move
the particle around, but this motion
is opposed by the friction
between a moving particle
and these same molecules.

The third term on the right-hand side contains the advection and phoresis terms.
Advection is motion of a particle because it is carried along
by the cytoplasm flowing at a local velocity ${\bf \vadvect}$.
${\bf U}$ is the diffusiophoretic velocity.
If the cytoplasm is inhomogeneous (has gradients) at a point, then locally the stresses
on the particle are also inhomogeneous, which means that there are unbalanced
stresses which will cause the particle to move relative to the local fluid \cite{And89,And86,Bra11,bocquet10,marbach17,yoshida17}.
This local motion is called a slip velocity, and can be caused by
gradients in anything. Here we will consider gradients in concentration,
and then this slip velocity
is a diffusiophoretic velocity, ${\bf U}$.
Both ${\bf U}$ and ${\bf \vadvect}$ are zero in a system at equilibrium, so in a cell
they must come from the cell's metabolism. 

The last term on the right is motion due to a 
force ${\bf f}$ on the particle, for example due
to a motor protein pushing or pulling on
the particle. In eukaryote cells, it is well established that motor proteins pull many cargos around the cell. Although this is an important process, it is well studied \cite{bionumbers_book} and so
here I only consider particles not being pulled by motor proteins.


To motivate this study, let us consider experimental evidence for metabolism-dependent mobility of particles in cells. Parry \etal~\cite{parry14}
studied the dynamics of large, around 50 to 150 nm across, particles in the
cytoplasm of bacteria
(including {\it E.~coli}). The particles included granules of an enzyme, a plasmid (of a type without an active partitioning system), and particles formed by a self-assembling viral protein. They found that the dynamics of particles
in this size range, dramatically slowed
down when the metabolism was
shut off. The metabolism was shut off by depleting ATP and GTP using 2,4-dinitrophenol (DNP).

Parry \etal~\cite{parry14} tracked the displacement of particles $\sim 100$~nm across over periods of 15~s. When the metabolism was shut down,
the particles made many fewer displacements of
order hundreds of nanometres,
and this dramatically slowed
movement.
So we are looking for a
metabolism-dependent mechanism that can
transport assemblies 100~nm across at an effective speed of up to $\sim 100$~nm/s for periods of 10~s.
Here I suggest that
diffusiophoresis is a possible mechanism.

It is worth noting that both with and without an active metabolism, the distribution of displacements was very far from the Gaussian distribution expected for thermal diffusion in a uniform background. This non-Gaussian distribution implies that the cytoplasm is strongly non-uniform.

The results of Parry \etal~\cite{parry14} are for bacteria. The presence of motors and the cytoskeleton in eukaryote cells, will make it difficult to unambiguously observe diffusiophoresis in eukaryotes. However, I note that Bajanca \etal~\cite{bajanca15} studied the motion of the protein dystrophin in the muscle cells of zebrafish embryos. This protein has been estimated to be 100~nm long. They found effective diffusion constants of order $1~\um^2$/s, only an order of magnitude lower than that of GFP ($\sim3$~nm across) in the same cells. The effective diffusion constant of dystrophin is seems too large to be consistent with the Stokes-Einstein expression for thermal diffusion, assuming an effective cytoplasmic viscosity ten times that of water. A cytoplasmic viscosity ten times that of water is consistent with the measured diffusion constant for GFP \cite{montero12}. This leaves us looking for a transport mechanism beyond simple thermal diffusion.


I am not the first to consider phoretic motion in cells,
Lipchinsky \cite{lipchinsky15} considered osmophoresis,
motion driven by a gradient in the osmotic pressure, in pollen tubes.
As the osmotic pressure gradient is due to a gradient in the concentration of small ions, osmophoresis is a type of diffusiophoresis.
 Ietswaart \etal ~\cite{ietswaart14}, Surovtsev \etal ~\cite{surovtsev16}, and Walter \etal ~\cite{walter17} all modelled what is called the ParA/B \cite{surovtsev18} system of segregating plasmid DNA in bacteria during cell division. The plasmid moves in a concentration gradient of the ParA protein, and so their work \cite{ietswaart14,surovtsev16,walter17} is an example of diffusiophoresis. However, the molecular interactions and stresses responsible for the plasmid motion were not explicitly modelled in that work \cite{ietswaart14,walter17}. Here I do consider these interactions and stresses here, and so my work is complementary to that earlier work \cite{ietswaart14,walter17}. Surovtsev \etal~\cite{surovtsev16} used a Brownian dynamics model for the interaction, this may overestimate the strength of diffusiophoresis, as discussed by Sear and Warren \cite{sear17,Bra11}.

\begin{figure}[b!]
 \includegraphics[width=8.0cm]{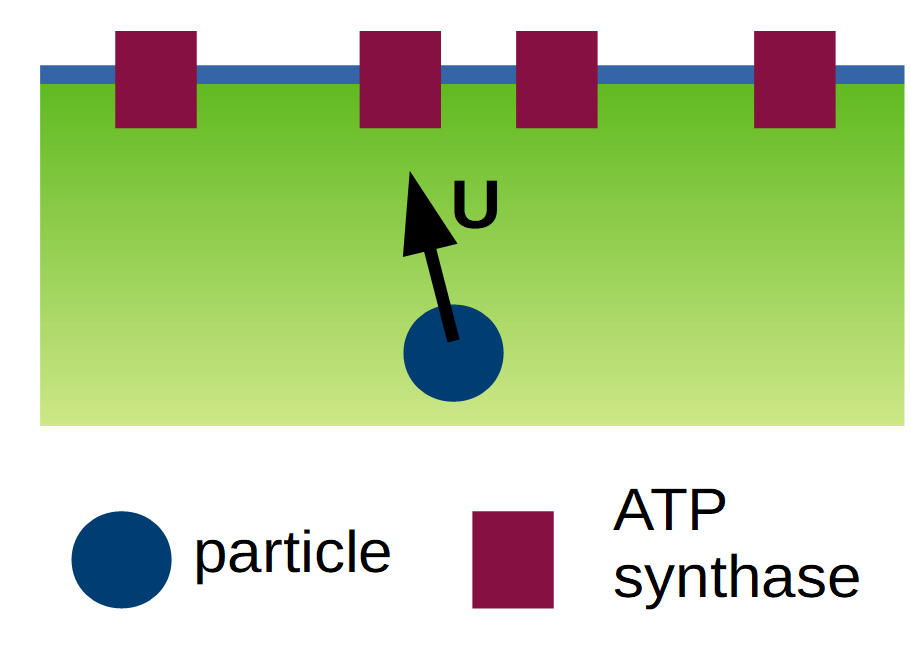}
\caption{
Schematic of part of a prokaryote cell, with an ATP concentration gradient indicated by shading.
Sources of the gradient are ATP synthases, in magenta, while our model
assumes that sinks (ATP consuming proteins) are uniformly distributed in the cytoplasm.
We show one particle moving up
the concentration at a diffusiophoretic velocity ${\bf U}$.
\label{cell_schem}}
\end{figure}

There are thousands of species inside cells, many of which may have gradients. To keep things simple, I work with the gradient in just one example species: the abundant metabolite ATP. I select ATP as a test candidate as it is known to interact strongly with proteins at the concentrations found in cells \cite{patel17}, and to turnover rapidly \cite{bionumbers_book}. The rapid turnover implies large fluxes between the sources and sinks, and the fluxes imply gradients, between these sources and sinks. Thus ATP is my best candidate for an abundant species whose concentration gradients I can estimate. When ATP is consumed ADP is produced, so although here I will refer to an ATP gradient for simplicity, in reality it is two gradients, one of ATP and one of ADP, with the opposite sense. The effects of these two opposing gradients may partially cancel, weakening diffusiophoresis, but as the molecules are different, any cancellation will be partial. Note that small ions
such as potassium and chloride are even more abundant than ATP inside cells, but as they do not turnover are expected to have only negligible concentration gradients. The numbers needed to
characterise cells in my
calculations are gathered together
in Table I in the Supplemental Material.
A particle moving up an ATP gradient is shown in \Figref{cell_schem}. 

Inside cells, thermal energy and momentum can move much more rapidly than even small molecules. So, I expect thermal and pressure gradients to be negligible, see the Supplemental Material for the justification of this assumption.





In order to estimate the sizes of the gradients in ATP concentration inside cells, I start by estimating the timescale for ATP to diffuse
across a typical bacterial cell $1~\um$ across. 
The diffusion constant of ATP both in water
and in cells \cite{bionumbers_book,graaf00}
is of order $100~\um^2$/s. So an ATP molecule diffuses
across the cell in of order 0.01~s.

An active $1~\um^3$ bacterial cell is estimated to have
$10^7$ ATP molecules and to consume $10^7$ ATP molecules each
second, see Table I of the Supplemental Material. This gives a 
time of 1~s between production by ATP synthase, and consumption. A lifetime 100 times the diffusion time implies gradients
of order 1\% to 10\% across a cell $1~\um$ across. For an ATP concentration of $10^7/\um^3$, we have gradients of $10^5/\um^4$ to $10^6/\um^4$.  I will use the gradient value $10^5/\um^4$ below.
See the Supplemental Material for a more detailed calculation that also gives gradients of this size. 
These are very simple estimates of steady-state gradients, the gradient will
presumably vary in space and time as particular sources
(ATP synthase) and sinks (ATP consuming proteins) move. But as ATP diffuses much faster than membrane proteins such as ATP synthase, ATP gradients may often be close to a steady state.



The diffusiophoretic velocity ${\bf U}$ is proportional to the gradient in the concentration $c$, of a solute
\begin{equation}
{\bf U}=\phorco\nabla c
\label{simple_daeq}
\end{equation}
There is a
standard Derjaguin/Anderson
expression \cite{ALP82,And89,Bra11,marbach17}
for the coefficient $\phorco$ that relates the concentration gradient to the diffusiophoretic velocity.
This expression is valid for a large particle with an interaction $\phi(z)$ between the particle surface and a smaller species that has a concentration gradient $\nabla c$.
Here $z$ is the distance separating the smaller species from the surface of the particle. Between the smaller species and the surface is
a continuum solvent with viscosity
$\eta$. The Derjaguin/Anderson
expression is
\begin{eqnarray}
\phorco &=&\frac{\kT}{\eta}\int_0^{\infty}
z\left[\exp\left(-\phi(z)/\kT\right)-1\right]
{\rm d}z  \label{eq:da}
\end{eqnarray}
Note that as the particle surface is interacting with the smaller species in water, $\phi(z)$ is an effective interaction free energy.

From \Eqref{eq:da}, we see that the diffusiophoretic coefficient $\phorco$ is approximately $\kT$ divided
by the solvent viscosity $\eta$, and multiplied by the
square of the interaction range, which we denote by $L$. So, we
obtain the approximate expression
\begin{equation}
 \phorco  \sim  \pm \kT L^2 / \eta
\label{simple_phorco}
\end{equation}
$\phorco$ is positive for
attractive interactions, and then
${\bf U}$ is directed to
higher concentrations of the solute. For repulsive interactions the sign is reversed.
The integral in \Eqref{eq:da} is of order $-L^2$ for a repulsive $\phi(z)$ that is $\sim \kT$ or stronger over a range $L$, and is of order $+L^2$ for an attractive $\phi(z)$ that is of order $\kT$ over a range $L$. For a stronger attraction, the integral will be larger, but \Eqref{eq:da} is an approximation \cite{ALP82,And89,Bra11,marbach17}, and will break down for strong enough attractions. To summarise, the approximation of \Eqref{simple_phorco} should be the correct order of magnitude unless there are attractions $\gg \kT$ in which case the Derjaguin/Anderson approximation fails. So I do need to assume that, for the particles studied by Parry \etal~\cite{parry14}, the interactions between the protein and ATP are not strongly ($\gg \kT$) attractive.


Here we estimate the
diffusiophoretic velocity ${\bf U}$ of a particle in a concentration gradient of ATP.
The diffusiophoretic coefficient depends on the free energy of particle/ATP interaction $\epsilon$,
the range of the surface/ATP interaction $L$, and the solvent viscosity $\eta$. I approximate the viscosity by that of water,
$\eta\sim 10^{-3}$Pa~s. The free energy of interaction $\epsilon$, I take to be $\kT=4\times 10^{-21}$J, and the range $L$ to be 1~nm.
From ATP's diffusion coefficient of $500~\um^2$/s, ATP has a Stokes-Einstein radius of $0.7$~nm.
Then $\phorco=4\times 10^{-18}\um^3$/s, and
\begin{equation}
U \sim 4\times 10^{-18}|\nabla \cATP |~~~~~[\nabla\cATP~~\mbox{in}~~\um^{-4}]
\label{umetabDP}
\end{equation}
for $\cATP$ the ATP concentration. We set $L=1$~nm, as that is the order of magnitude of both the size of ATP itself and of the Debye screening length in the cytoplasm. ATP is both highly charged and contains organic groups, so its nature is a little amphiphilic. Therefore, the interactions with a protein surface will be complex \cite{patel17} but will include electrostatic interactions, with a range of the Debye length. Interactions beyond a few nanometres are expected to be weak \cite{israelachvili_book}.

Above, we estimated the gradient in ATP concentration to be $10^{5}/\um^4=10^{29}/$m$^4$.
Putting that gradient in \Eqref{umetabDP}, we have a
diffusiophoretic speed $U\sim 400$~nm/s. This is large enough
to be consistent with the motion observed by Parry \etal~\cite{parry14}, so long as the gradient lasts for of order 10~s or more. Our estimates for the gradients, are steady-state estimates, so they should satisfy this constraint.

This is the key result of this work: Physically reasonable concentration gradients of one abundant metabolite, can drive
motion of large particles that is fast
enough to be significant for transport inside cells, and fast enough to be observable.
Note that as typical proteins
diffuse across a $1~\um$ cell
in less than 1 second, an additional
speed of 100~nm/s has little
effect on the dynamics of single proteins, so diffusiophoresis should not affect significantly affect protein dynamics.

My estimate of speeds of hundreds of nanometres per second is highly approximate, so I would like to comment on sources of uncertainty. It relies on my estimate of the gradients. These could be out by an order of magnitude, and it is difficult to assess how gradients vary in space and time. It is also worth noting if the phoretic velocity is directed towards a source of a gradient, there will be positive feedback as particles will be pulled towards the source where the gradient is steepest, an effect that is magnified when the source itself can move \cite{reigh18}.
The phoretic interaction could
pull the particle into contact with the source, where the concentration gradients are strongest.
This was a theory and simulation study. Experiments {\em in vitro} by Zhao \etal ~\cite{zhao18} found that phoretic interactions can help enzymes move together. Thus our estimates for $U$ may
be underestimates when the phoretic
velocity is towards gradient sources.

The estimated speed also relies on our value for $\phorco$. The Anderson-Derjaguin expression \cite{ALP82,And89,Bra11,marbach17} applies to dilute systems (the cytoplasm is not dilute), and relies on flow in a fluid interfacial region of width $L$, driven by the stresses there. It is uncertain how good these approximations are in the cytoplasm.

There have been ({\it in vitro}) experimental studies of proteins moving due to active processes.
Sen and coworkers \cite{zhao17,zhao18,zhao18b,mohajerani18}, and Granick and coworkers \cite{jee18}, have both studied enzymes, such as urease, in dilute solution.
Both groups find that enzymes
move faster when they are catalysing
reactions, and Zhao \etal ~\cite{zhao17} also found that active enzymes could speed up the motion of other species. Future work could consider solutions with concentrations of energy-consuming molecules that are closer to those found in the cytoplasm. Jee \etal ~\cite{jee18} have already considered the effect of a crowding agent. Future work could also use microfluidics to create gradients in ATP, in order to look for phoresis.

My estimate is for prokayotes.
Milo \etal~\cite{bionumbers_book} discuss
the energy consumption of mammalian
cells. The power consumption per unit
volume of a fibroblast can be comparable
to that of {\em E.~coli}. Assuming
distances of a few
micrometres between where ATP is consumed,
and mitochondria, the ATP gradients in an active
fibroblast will be comparable to those in
growing {\em E.~coli}. So diffusiophoretic speeds should also be comparable.

We have only considered a gradient in one of the thousands of species in a cell (ATP), and models of the ParA/B system of moving plasmids in bacteria \cite{walter17,surovtsev16,surovtsev18} also only consider ParA gradients. Future work will need to deal with the multicomponent nature of the cytoplasm. Systems that have evolved to localise species such as plasmids presumably have to work against the forces due to flutuating gradients in the other species present in the cell.


Diffusiophoresis is unlikely to be the only non-motor-driven metabolism-dependent transport mechanism in cells.
See the Supplemental Material for more discussion of these other potential transport mechanisms. In eukaryote cells, there is also transport of particles as the cargos of motor proteins.




In conclusion, the more we learn of the cytoplasm of both prokaryote and eukaryote cells, the less uniform they appear to us \cite{luby99,agutter00,shin17cliff,woodruff17}. There must be many gradients in cells, and so phoresis
must be occurring in essentially all
cells. However, quantifying phoretic
speeds in cells is difficult.
Cells are complex, and the size of gradients is unknown. In addition the interactions needed to estimate
diffusiophoretic coefficients $\phorco$ are
also unknown. Here I estimated
$\phorco$ for ATP, and estimated
the size of gradients of
ATP in an active bacterial
cell such as {\em E.~coli}.
I predicted that diffusiophoretic speeds
of order 100~nm/s are possible. This is large enough
to be consistent with the motions
observed by Parry \etal ~\cite{parry14},
for large (50 to 150~nm) particles. However, the complexity of the cytoplasm means that is very difficult to unambiguously show that observed movements are due to one specific transport mechanism.
Experiments on simpler, {\em in vitro}, systems will probably be required
to separate out different non-thermal-diffusion contributions to transport in cells.

\begin{acknowledgments}
I would like
to thank Patrick Warren for teaching me
much of what I know about
diffusiophoresis, and Daan Frenkel
for many illuminating discussions.
I would also like
to thank
the organisers, Julian
Shillcock, Mikko Haataja and John Ipsen, and participants of the CECAM workshop
{\em Liquid Liquid Phase Separation in Cells}, for helpful questions and feedback.
The author confirms that no new data were created
during this study.
\end{acknowledgments}

\section{{\large SUPPLEMENTAL MATERIAL}}

\begin{table}[bth!]
\centering
 \begin{tabular}{|c | c|} 
  \hline
bacteria cell size & $1~\um$\\
\hline
total protein concentration \cite{bionumbers_book} & 
$c_{{\mathrm P}}\simeq 3\times 10^6/\um^3$\\
in cytoplasm & $c_{{\mathrm P}}\simeq 3\times 10^6$/cell \\
\hline
total metabolite concentration \cite{bionumbers_book,bennett09} & 
$c_{{\mathrm M}}\simeq 10^8/\um^3$ \\
in cytoplasm & $c_{{\mathrm M}}\simeq 200$~mM \\
 & $10^8$/cell \\
\hline
ATP concentration \cite{bionumbers_book,bennett09}
& $c_{{\mathrm ATP}}\sim 10^{7}$/$\um^3$ \\
 in cytoplasm & $10^{7}$/cell \\
 \hline
ATP diffusion constant \cite{hubley96,graaf00} & $D_{\mathrm{ATP}}\sim 5\times 10^{2}\um^2$/s \\
 (in solution \& in vivo) &  \\
\hline
ATP hydrodynamic diameter \cite{bionumbers_book} & 1.4 nm \\
\hline
ATP molecular weight & 507 g/mol \\
\hline
Debye length $\kappa^{-1}$ in 200 mM KCl & $0.7$~nm \\
\hline
viscosity of water & $\eta_W\sim 10^{-3}$Pa~s\\
 \hline
Power consumption & $P\sim 10^{-12}$W\\
of bacterial cell\cite{bionumbers_book,jain09} &
$P\sim 10^7$ATP/s
\\
\hline
 \end{tabular}
 \caption{Table of the values I use
 for properties of the cytoplasm.
The value for the concentration of ATP is from Bennett \etal ~\cite{bennett09}, who give 10~mM as the ATP concentration in their rapidly growing {\em E.~coli}. Other measurements \cite{bionumbers_book} for {\em E.~coli} have given similar but typically a little lower values, and Traut \cite{traut94} gives
3~mM~$\sim 10^6/\um^3$ for example mammalian cells. The concentration of small cations (eg K$^+$ and Na$^+$) and anions (eg Cl$^-$) in {\it E.~coli} will vary with external growth conditions, but 200 mM is a typical value \cite{bionumbers_book}.
N.B., 1 mM~$\simeq 6\times 10^5/\um^3$.
Most of the other numbers are from
{\em Cell Biology by the Numbers} \cite{bionumbers_book,bionumbers_web}.
}
 \label{table1}
\end{table}

\section{Numbers for properties of a bacterial cell}
\label{app_numbers}

To estimate diffusiophoretic velocities
in a bacterial cell, we need estimates
for a number of properties
of a typical bacterial cell, by which I mean {\em E.~coli}. These are collected in
Table \ref{table1}.

Many of these numbers are from
the excellent reference,
{\em Cell Biology by the Numbers} by Milo and Phillips. This is available as both
a book \cite{bionumbers_book},
and an online resource
\cite{bionumbers_web}.

\section{Estimation of the size of gradients in the temperature}
\label{app_thermo}

The relative rates of heat and molecular diffusion, is characterised by the Lewis
number: Le$=\alpha/D$. Here $\alpha$ and $D$ are the thermal diffusivity, and the diffusion constant
of the molecule, respectively. Even for fast diffusing species,
such as molecules like ATP, the diffusion constant
$D\sim 100~\um^2$/s \cite{graaf00,bionumbers_book},
while for water (and hence the cytoplasm
which is mainly water)
$\alpha\sim 10^{5}\um^2$/s \cite{ramires95}.
Thus for small molecules in the
cytoplasm, $\mathrm{Le}\sim 10^3$,
and so temperature gradients relax
about a thousand times faster than gradients in ATP.

Momentum diffuses with a diffusion constant of the kinematic viscosity, about $\nu\sim 10^{6}\um^2$/s for water. Thus in cells pressure gradients relax even faster than temperature gradients, and we therefore expect pressure gradients to be negligible.

We can estimate the size of temperature gradients as follows. A growing bacterial cell of volume
$1~\um^3$ has a power consumption of order $10^{-12}$W, see Table \ref{table1}. If
we naively assume that the metabolism
is concentrated in say, the
left-half of the cell, then
crossing the midpoint of the cell
we have of order $10^{-12}$W of heat,
or $\sim 1$~W/m$^2$, for a cell
cross-section of $1~\um^2$.
The thermal conductivity of water
is of order 1~W/K/m \cite{ramires95},
so a flux of 1~W/m$^2$ is driven
by a gradient of order 1~K/m.

Thus we conclude that active bacterial
cells have temperature gradients across
of them that are of order 1~K/m, or that
the temperature differences across the cells
are no more than 1~$\upmu$K.
This is a general observation, cells
are made of matter with high thermal
conductivity, and so cannot
support significant temperature
gradients. So, presumably,
the recent claim \cite{chretien18}
that mitochondria
are 10~K hotter than the surrounding
cytoplasm is incorrect.

\section{Gradient in ATP across the cytoplasm}
\label{app:gradlength}

To obtain a simple estimate of the size of ATP gradients I assume a one-dimensional geometry in which ATP
synthases are along two parallel flat cell walls at $z=\pm w/2$. I assume that the system is at steady state. As the time taken for ATP to diffuse across the cell is only 0.01 s, steady state will be achieved in much less than a second.
The cell width $w=1~\um$. To get a simple one-dimensional model I then ignore gradients parallel to the wall,
and assume the concentration of ATP depends only on the distance $z$ from
the wall. If the proteins consuming
ATP (= the ATP sinks) are uniformly distributed, then
the concentration of ATP in the cytoplasm obeys
\begin{equation}
\DATP\left(\frac{{\rm d}^2\cATP(z)}{{\rm d} z^2}\right)
-k_{\mathrm{ATP}}\cATP(z)=0
\end{equation}
Here $\DATP$ is the diffusion constant for ATP, and $k_{\mathrm{ATP}}$ is the rate constant for ATP consumption, assumed uniform in the cytoplasm.
If the ATP synthases along the cell wall maintain the ATP concentration
at a fixed value $\cATP(z=\pm w/2)$, this provides the boundary conditions needed to solve
this differential equation. The solution is then
\begin{equation}
\cATP(z)=\cATP(z=\pm w/2)
\frac{\cosh\left(-z/\gradlength\right)}{\cosh\left(w/\left(2\gradlength\right)\right)}
\end{equation}
with the lengthscale of the gradient $\gradlength=\left(\DATP/k_{\mathrm{ATP}}\right)^{1/2}$.
The gradients are then of order
$\cATP(z=\pm w/2)/\gradlength$.
For $\DATP= 100~\um^2/$s and $k_{\mathrm{ATP}}=1$/s,
$\gradlength=10~\um$. Using $\cATP(z=\pm w/2)=10^7/\um^3$, the
gradients $|\nabla\cATP|\sim 10^6/\um^4$.
This one-dimensional model neglects
both the discrete nature of the ATP
source (ATP synthase at the membrane),
and fluctuations.



\section{Gradients of metabolites near a metabolon}
\label{app:metabolon}

In the main part of this paper we considered one metabolite: ATP. Here we consider a metabolite produced/consumed by a large protein complex --- these large complexes are sometimes called metabolons \cite{sweetlove18}. 
Metabolons are physical assemblies
of many proteins, including copies
of multiple species of enzyme in
the same pathway, \ie~if a synthetic pathway requires enzymes
A, B and C, with B catalysing a reaction
on a product of enzyme A, etc, then
many of the copies of A, B and C
may be together in an physical assembly
of perhaps hundreds or thousands
of molecules. This
may enhance the efficiency of this
pathway \cite{sweetlove18}.
I will show that near these metabolons, we should also expect large gradients in the concentrations of metabolites.

For simplicity, I approximate a metabolon by
a sphere of radius $R_{\mathrm{MN}}$,
producing fluxes of order $k=10^5$/s, of a single
molecule with diffusion constant $D_{\mathrm{M}}$.
A single urease can catalyse the hydrolysis of urea at a rate of $10^3$/s \cite{jee18}, so this flux could be produced by a hundred copies of a high turnover enzyme. I assume that just the reactants interact with the particle, including products just
complicates the expressions a little.

Our model is essentially that studied
in detail by Reigh \etal ~\cite{reigh18}.
Following Reigh \etal, we estimate
the steady-state gradient.
At steady state, the
concentration $c_{\mathrm{M}}$ obeys Laplace's equation
$\nabla^2c_{\mathrm{M}}=0$.
I use spherical coordinates
centred on the metabolon,
with $r$ the distance from the centre
of the metabolon. Then the flux is 
\begin{equation}
\nabla c_{\mathrm{M}}(r)=-\frac{k}{4\pi D_{\mathrm{M}} r^2}\hat{r}
\end{equation}
as this gives the required total
flux $k$ over a surface enclosing the metabolon.
For a metabolite
diffusion coefficient
$D_{\mathrm{M}} \sim 100~\um^2$/s, and $k=10^5$/s, the flux is
\begin{equation}
\nabla c_{\mathrm{M}}(r)\sim-\frac{10^{14}}{r^2}\hat{r}
\end{equation}
where we approximated $4\pi$ by 10, as the expression is approximate.
At a distance of order 100~nm from the centre of the metabolon, the gradient is of order
$10^{28}$/m$^4$ or $10^{4}/\um^4$.

Reigh \etal ~\cite{reigh18} tested the simple theory above by
essentially exact computer simulations of a simple model. There was semiquantitative agreement between the theory and computer simulations.



\section{Gradients of small ions such as potassium and chloride}

Small ions such as potassium, sodium and chloride are abundant in cells, $10^8/\um^3\sim 100$~mM \cite{bionumbers_book}, but we expect the gradients in their concentration to be very small. So we do not expect significant phoretic effects due to gradients in the concentration of small ions, in cells growing in an environment where the osmotic pressure is constant.

The timescale for potassium turnover in {\it E.~coli} has been measured at of order $10^3$s, \cite{schultz62}. Potassium is the most abundant cation in cytoplasm, while chloride is the most abundant anion \cite{bionumbers_book}. Presumably, due to electroneutrality, the flux of anions and cations has to be the same.

The diffusion constant of potassium chloride in water is of order $10^{3}\um^2$/s \cite{gosting50}, so a potassium ion will diffuse across a bacterial cell in about 1 ms. This is a factor of $10^6$ times smaller than the timescale for potassium uptake, and so we expect the gradients in cells, of the concentration of potassium, and chloride, to be very small. As diffusiophoresis is driven by gradients, this implies that diffusiophoresis driven by small ions should typically be irrelevant. An exception may be pollen tubes, a very specialised and large type of cell where large gradients are found, see the work of Lipchinsky \cite{lipchinsky15}.

\section{Alternative metabolism-dependent mechanisms of transport in cells}
\label{app:alternate}

Diffusiophoresis is unlikely to be the only non-motor-driven metabolism-dependent transport mechanism in cells.
In this section, I briefly consider two other possible mechanisms for transport in cells, that rely on the cell's metabolism. These are advection of a particle due to flow in the cytoplasm, and metabolism-dependent processes accelerating thermal diffusion by making the cytoplasm less sticky. In eukaryote cells, there is also transport of particles as the cargos of motor proteins.

\subsection{Transport by flow}

\subsubsection{Cytoplasmic streaming}

It is clear that in a number of large cells ($\gtrsim 100~\um$), there is significant flow of the
cytoplasm. This corresponds to a large $\vadvect$ term in Eq.~(1) in the main text, with a $\vadvect$ that is relatively uniform over large regions of the space, and relatively constant in time,
This is sometimes called cytoplasmic streaming \cite{goldstein15,ganguly12}. Cytoplasmic streaming is driven by motors and the cytoskeleton,
and it clearly contributes to transport in a number of very large cells \cite{goldstein15,ganguly12}.
These large cells include $100~\um$ {\em Drosophila} oocytes \cite{ganguly12}, and plant cells that can be centimetres
long \cite{goldstein15}.
Speeds of tens of nanometres per second
were measured in the oocytes, while
much faster speeds are found in larger cells.
I am not aware
of studies of cytoplasmic streaming
in eukaryote cells of more typical
size, $\sim 20~\um$ across, or in prokaryote cells.

\subsubsection{Random stirring of the cytoplasm}

Mikhailov and Kapral \cite{mikhailov15,kapral16} have considered stirring of the cytoplasm by energy-consuming but non-motor proteins.
Here by stir, I mean generate transient flow in more-or-less random directions in the cytoplasm, as opposed to the fast directed flow seen in cytoplasmic streaming. So, here $\vadvect$ varies rapidly in space and time. They considered proteins that consume ATP and generate force dipoles, which stir the surrounding cytoplasm, thus accelerating diffusion in this cytoplasm. Note that proteins free in the cytoplasm generate force dipoles not forces, due to Newton's Third Law.

They studied active proteins at a concentration $c_{\mathrm{FD}}$, that  generate force dipoles of root-mean-strength strength $F\delta$ with
a characteristic correlation time $\tau_{\mathrm{FD}}$. Mikhailov and Kapral found that these force dipoles increase diffusion by an amount (Eq.~(10) of Mikhailov and Kapral \cite{mikhailov15})
$\Delta=0.01c_{\mathrm{FD}}(F\delta)^2\tau_{\mathrm{FD}}/\eta^2l_{\mathrm{c}}$, for $l_{\mathrm{c}}$ a small lengthscale cutoff,
approximately equal to the distance of closest approach between the
active protein, and the protein whose diffusion is being accelerated.
This follows from the
fact that a force dipole
induces flow
at speed $v\sim F\delta/\eta r^2$, a distance $r$ away.

Subsequent computer simulations
of a simple model system, by Dennison
\etal ~\cite{dennison17} found
a relatively small effect on diffusion,
of order 10\% or less. However, the size of the increase in diffusion is very sensitive to a number
of parameters so it is hard to estimate how large an affect it could have in cells, without better data on the cytoplasm.

\subsection{Metabolism-dependent viscosity}

The speed of diffusion is reduced by drag on the diffusing particle.
This drag will increase if the particle sticks to the proteins
in the cytoplasm. Thus any energy-consuming process, such as those
involving chaperones, which unsticks proteins, will cause a
metabolism-dependent increase in diffusion. We do not know if
such a process contributes to the results of Parry \etal ~\cite{parry14},
or is a general source of metabolism-dependent diffusion.

It is also worth noting that the metabolism is typically inhibited by
depleting the ATP in a cell. ATP at physiological
concentrations, is known \cite{patel17} to strongly interact with proteins. Recent work of Patel \etal ~\cite{patel17}
 showed that ATP inhibits proteins undergoing liquid/liquid
phase separation. The liquid/liquid separation is into coexisting
phases with high and low concentrations of
protein. Note that this effect of ATP is due
to physical interactions between ATP (a relatively large and amphiphilic ion)
and proteins, the ATP is not consumed, it is a purely equilibrium effect.
Thus when depleting the
ATP in a cell, some interactions of the particle may change, in addition to the suppression of the metabolism removing ATP gradients.


%

\end{document}